\documentstyle[numreferences]{jobp}  

\begin{opening}
\title{Quantum {\upc G}roup {\upc S}ymmetries in {\upc C}onformal
{\upc F}ield {\upc T}heory}
\author{Krzysztof Gawedzki}
\institute{CNRS, IHES, 91440 Bures-sur-Yvette, France}
\date{}
\end{opening}

\begin{document}
\begin{abstract}
\hspace{-0.2cm}{}\footnote{Talk given at Oji
Seminar on Quantum Analysis, Kyoto, June 25-29,
1992}\hspace{0.2cm}
Quantum groups play the role of hidden symmetries of some two-dimensional
field theories. We discuss how they appear in this role
in the Wess-Zumino-Witten model of conformal field theory.
\end{abstract}

\section{WZW model}



\ \ \ Wess-Zumino-Witten (WZW) model [1] occupies an important place in
the conformal field theory: it is a model of two-dimensional
quantum massless fields exactly soluble for energy spectrum and
correlation functions; it is also a generating theory for
a rich family of other conformal field theories which may be
obtained from the WZW one by various versions of the so called
coset construction [2][3]. It is a $1+1$-dimensional analogue of
$0+1$-dimensional particle on the group $G$ (for simplicity,
we shall assume $G$ to be a compact matrix group). The motion
of the particle on $G$ is described by the classical equation
$$\partial_t(g\partial_tg^{-1})=0.\eqno(1)$$
On the quantum level, the system may be solved by harmonic analysis
on $G$. The quantum space of states is $L^2(G)$ and
carries the representation of $G\times G$.
$$L^2(G)=\bigoplus\limits_\lambda V_\lambda
\otimes{\overline V_\lambda}\eqno(2)$$
where $V_\lambda$ denotes the irreducible highest weight (HW) $\lambda$
representation of $G$. The Hamiltonian is proportional to the
quadratic Casimir operator. The multiplication
by matrix elements of $g\in G$ may be expressed in the realization
(2) of $L^2(G)$ as a bilinear combination of Clebsch-Gordan
coefficients for the tensor product with the fundamental
representation, i.e. as a bilinear expression in intertwiners of
representations of $G$.
\vskip 0.2cm

The WZW model may be viewed as describing the particle on
the loop group $LG$. The classical equation of motion is
$$\partial_-(g\partial_+g^{-1})=0\eqno(3)$$
where $\partial_\pm\equiv\partial/\partial(x^1\pm x^0)$
and we shall consider the cylindrical geometry with $x^1$ taken
mod $2\pi$.
The quantum theory is solved by harmonic analysis on the
Kac-Moody group $\widehat{LG}$, the central extension
of the loop group.  The space of states is
$$\bigoplus\limits_{\lambda\ {\rm int.}}
V_{k,\lambda}\otimes{\overline V_{k,\lambda}}
\eqno(4)$$
where the level $k$, a fixed positive integer, is the value of the
central charge of $\widehat{LG}$ and $\lambda$ runs through the
finite ($k$-dependent) set of the so called integrable highest weights.
The symmetry of the model is $\widehat{LG}\times\widehat{LG}$ where the
factors act on, respectively, left- and
right-moving degrees of freedom. The spaces $V_{k,\lambda}$
and $\overline V_{k,\lambda}$ combine diagonally in eq. (4)
due ot the coupling between the left- and right-movers which share
a finite number of degrees of freedom. The field operators of the
WZW theory are bilinear combinations of intertwiners of
representations of $\widehat{LG}$.
\vskip 0.2cm

There are several ways to see manifestations of the hidden
quantum group symmetry in the WZW model\footnote{See also [4]-[6]
for the related work concerning the minimal or Liouville theories
and [7]-[9] for an attempt at a more general, less model dependent
approach relating the quantum-group-like symmetries to the theory of
superselection sectors and field statistics in the
spirit of [10][11]}. The labeling of the
representations $V_{k,\lambda}$ is similar to that of the
``good'' representations [12]  of the Drinfeld-Jimbo quantum
group [13][14] $\hspace{0.05cm}{\cal U}_q(\cal G)$ for $q={\rm e}^{\pi
i/(k+h{\check{}
\hspace{0.05cm}})}$
($h{\check{}\hspace{0.05cm}}$ is the dual Coxeter number of $\cal G$,
the Lie algebra of $G$).
The exchange algebra of the intertwining operators is given
by the $6j$-symbols of ${\cal U}_q(\cal G)$ [15]-[17]. In the free
field description of the WZW theory [18][19], the quantum group
describes the homology of the contours of screening charge integrals
[20]-[23]. Let $j(x)$ be the current giving the infinitesimal
version of a representation $\pi$ of $\widehat{LG}$ by
$$\pi(X)=\int\limits_0^{2\pi}{\rm tr}\hspace{0.05cm}j(s)\hspace{0.05cm}
X(s)\hspace{0.05cm}ds\eqno(5)$$
for $X\in L{\cal G}$. In representation of central charge $k$,
current $j$ satisfies the commutation relations
$$[j(x)_1,j(y)_2]=(j(x)_1C-Cj(y)_2)\hspace{0.05cm}\delta(x-y)
-(ik/2\pi)\hspace{0.05cm}C\hspace{0.05cm}\delta'(x-y)\eqno(6)$$
where $\hspace{0.05cm}j_1\equiv j\otimes Id$\hspace{0.05cm},
$\hspace{0.05cm}j_2\equiv Id\otimes j\hspace{0.05cm}$ and
$\hspace{0.05cm}C=\sum t^a\otimes t^a\in{\cal G}\otimes{\cal G}
\hspace{0.05cm}$, $\hspace{0.08cm}{\rm tr}\hspace{0.08cm}
t^at^b=\delta^{ab}\hspace{0.05cm}$.
Eq. (6) may be viewed as the defining relation of the Kac-Moody
algebra $\widehat{L{\cal G}}$. Various aspects of the
$\hspace{0.05cm}{\cal U}_q(\cal G)$ symmetry in the
WZW model may be formally explained by postulating that the monodromy
operators
$$M(x)=P\hspace{0.07cm}{\rm e}^{\int\limits_x^{x+2\pi}j(s)
\hspace{0.07cm}ds}\eqno(7)$$
satisfy the $\hspace{0.05cm}{\cal U}_q(\cal G)$ relations
$$M_1R^-M_2(R^+)^{-1}=R^-M_2(R^+)^{-1}M_1\eqno(8)$$
where $R^\pm$ is a pair of solutions of the quantum Yang-Baxter
equation
$$R^\pm_{12}R^\pm_{13}R^\pm_{23}=R^\pm_{23}R^\pm_{13}R^\pm_{12}.\eqno(9)$$
Operators (7) appeared (implicitly) in [24] and were explicitly
discussed as generators of $\hspace{0.05cm}{\cal U}_q(\cal G)$
in [25]. The problem is that, as written above, they are very singular
objects due to the short distance singularity of currents $j(x)$.

\section{Lattice Kac-Moody algebra}

\ \ \ A possible way out from the above difficulty has been proposed
in a series of papers [26]-[29] which introduced the lattice version
of the Kac-Moody algebra\footnote{See [30]-[33] for related proposals
concerning the Liouville and Toda theories}. This allows
to build the regularized version
of the WZW model preserving essentially all the symmetries of the
continuum WZW theory and at the same time making explicit the quantum
group symmetry hidden in the continuum version of the theory.
\vskip 0.2cm

View ${\bf Z}_N$ as an $N$-point lattice in $S^1$.
The lattice Kac-Moody algebra ${\cal K}_N$ proposed in [26][27]
is given by matrix generators $J(n)$, $n\in {\bf Z}_N$, satisfying
the quadratic relations
\begin{eqnarray*}
\hbox to 3cm{$J(n)_1J(n)_2$\hfill}&=&R^+J(n)_2J(n)_1R^-,\\
\hbox to 3cm{$J(n)_1R^-J(n-1)_2$\hfill}&=&J(n-1)_2J(n)_1,\\
\hbox to 3cm{$J(n)_1J(m)_2$\hfill}&=&J(m)_2J(n)_1\ \ \
{\rm for}\ \ |n-m|>1.
\end{eqnarray*}
The other relations are the deformations of the $det=1$ conditions
for $SL(n)$ etc. For $SL(2)$ they read
$J_{11}(n)J_{22}(n)-q^{-1}J_{21}(n)J_{12}(n)=q^{1/2}$
(here the subscripts ``ij'' refer to the matrix element of $J(n)$).
The $SL(2)$ $R$-matrices are
$$R^+ = q^{1/2}\left(\matrix{q^{-1}&0&0&0\cr 0&1&q^{-1}-q&0\cr
0&0&1&0\cr 0&0&0&q^{-1}}\right) ,\hspace{0.1cm}R^- =
q^{-1/2}\left(\matrix{q&0&0&0\cr
0&1&0&0\cr 0&q-q^{-1}&1&0\cr 0&0&0&q}\right).$$
Operators $J(n)$ should be thought of as regularized versions
of parallel transport operators $\ P\hspace{0.07cm}\exp[{(k+
h\check{}\hspace{0.05cm})^{-1}\int_{2\pi n/N}
^{2\pi(n+1)/N}j(s)\hspace{0.07cm}ds}]$\hspace{0.2cm}[26].
\vskip 0.2cm

Algebra ${\cal K}_N$ contains quantum group $\hspace{0.05cm}
{\cal U}_q({\cal G})\hspace{0.05cm}$ as the monodromy
which is now, unlike in eq. (7), defined by a regular expression:
$$M(n)=J(n+N-1)J(n+N-2)\cdots J(n).\eqno(10)$$
$M(n)$ satisfy relations (8). For the $SL(2)$ case, we may write
$$M=q^{3/2}\left(\matrix{q^{-S^3}&(q^{-1}-q)S^+\cr 0&q^{S^3}}\right)
\left(\matrix{q^{S^3}&0\cr (q-q^{-1})S^-&q^{-S^3}}\right)\eqno(11)$$
where
\begin{eqnarray*}
[S^3 ,\hspace{0.05cm}S^\pm] &=& \pm S^\pm ,\cr
[S^+ , S^{-} ] &=& (q^{2S^3}-q^{-2S^3})/(q-q^{-1}).
\end{eqnarray*}
Algebras ${\cal K}_N$ may be also viewed as interpolating
between quantum group ${\cal U}_q({\cal G})$ and the
enveloping algebra ${\cal U}(\widehat{L{\cal G}})$ when
$N$ goes from $1$ to $\infty$.

\section{Conformal invariance on lattice}

\ \ \ The algebraic manifestation of the conformal invariance
underlying the continuum Kac-Moody algebras is the action
of the orientation-preserving diffeomorphisms
of the circle $D\in{\rm Diff}_+S^1$ on the currents
$$j(x)\mapsto {_{dD(x)}\over{^{dx}}}\hspace{0.05cm}j(D(x))$$
which induces automorphisms of $\widehat{L{\cal G}}$. The latter
are unitarily implementable in the HW representations:
$${_{dD(x)}\over{^{dx}}}\hspace{0.05cm}j(D(x))=U_D\hspace{0.08cm}
j(x)\hspace{0.08cm}{U_D}^{-1}.$$
$D\mapsto U_D$ is a projective representation of ${\rm Diff}_+S^1$
giving, on the infinitesimal level, a representation of the Virasoro
algebra.
\vskip 0.2cm

This structure essentially descends to the lattice. Let
$$D:{\bf Z}\rightarrow{\bf Z}$$
be an increasing map such that $D(n+N')=D(x)+N$ ($N'\leq N$).
Thus $D$ describes blocking of intervales of lattice ${\bf Z}_N$
into those of ${\bf Z}_{N'}$. $D$ induces a ``block spin''
homomorphism of the lattice Kac-Moody algebras
${\cal D}:{\cal K_{N'}}\rightarrow{\cal K_N}$
$${\cal D}(J(n))=J(D(n+1)-1)\cdots J(D(n)+1)J(D(n)).\eqno(12)$$
{\bf On the lattice, the conformal transformations are represented
by a (local) renormalization group!}
\vskip 0.2cm

If we relabel algebras ${\cal K}_N$ by arbitrary triangulations $T$
of $S^1$ (i.e. splittings of $S^1$ into intervals $l_n$)
renaming the generators as $J(l_n)$ then the block spin homomorphisms
give rise to homomorphisms $\iota_{T',T}:{\cal K}_T\rightarrow
{\cal K}_{T'}$ for $T'$ finer than $T$. One may define
the continuum limit algebra\footnote{Its generators are
$\ P\hspace{0.07cm}\exp[\int_x^y(k+h\hspace{0.04cm}\check{})^{-1}
j(s)\hspace{0.05cm}ds]$\ \ which
are singular in the standard continuum Kac-Moody algebra}
${\cal K}_{\infty}$ as the inductive
limit of algebras ${\cal K}_T$. Any $D\in{\rm Diff}_+S^1$ defines
an isomorphism ${\cal D}:{\cal K}_T\rightarrow{\cal K}_{D(T)}$
which descends to ${\cal K}_\infty$. As a result, ${\rm Diff}_+S^1$
acts by automorphisms on ${\cal K}_\infty$ showing that the block
spin homomorphisms indeed encode the conformal invariance.
\vskip 0.2cm

In continuum, the generators of the Virasoro
algebra implementing conformal invariance in the HW representations
of the Kac-Moody algebra are given in terms of current $j(x)$
by the Sugawara construction. On the lattice, the block spin
homomorphisms $\cal D$ may be also implemented in the class of
representations that we shall study below. If $D$ is a rigid
rotation of ${\bf Z}_N$ then we may expect that the implementing
maps are expressible in terms of generators of ${\cal K}_N$.
Such lattice Sugawara expressions for the (Minkowski time) transfer
matrix are not known yet except for the $U(1)$ case, see below.

\section{Free field representations of the Kac-Moody algebras}

\ \ \ In continuum, there are various ways to approach the construction
of the HW representations $V_{k,\lambda}$ of the Kac-Moody algebras.
\begin{enumerate}
\item In the algebraic approach [34] one starts from the concept
of Verma modules of $\widehat{L{\cal G}}$ and analyzes
their reducibility studying the structure
of singular vectors with the Kac-Kazhdan determinant formula,
constructing the BGG resolution etc.
\item In the geometric Borel-Weil type approach [35] one constructs
$V_{k,\lambda}$ as the space of holomorphic sections of a line bundle
over $LG/T$ ($T$ is the Cartan subgroup of $G$).
\item Finally, more recently, another algebraic approach to
the HW representations of $\widehat{L{\cal G}}$ has been obtained
[18][19][36][37] by representing
current $j(x)$ by free fields and constructing
$V_{k,\lambda}$ as a cohomology of a complex of Fock space (Wakimoto)
modules of the Kac-Moody algebra.
\end{enumerate}
\vskip 0.2cm

{}From those three approaches at least the last one carries over
to the lattice so let us sketch some of its essential points.
We shall stick to the $SL(2)$ case. The free fields one uses
to represent $j(x)$ are chiral scalar field $\phi(x)$
satisfying
$$[\phi(x),\phi(y)]={_{\pi i}\over^{2(k+2)}}\hspace{0.07cm}{\rm sgn}
(x-y)\eqno(13)$$
or the corresponding
$u(1)$ current $\partial\phi$ \hspace{0.05cm}(\hspace{0.03cm}
$\phi(x+2\pi)=\phi(x)+\pi r/(k+2)$
where $r$ is the momentum) and the $\beta\gamma$ system
$\beta(x)$, $\gamma(x)$ (periodic in $x$) s.t.
$$[\beta(x),\gamma(y)]={_{2\pi i}\over^{k+2}}\hspace{0.07cm}\delta(x-y)
\eqno(14)$$
(all other commutators vanish). These fields act in the standard Fock spaces
${\cal F}_r$ (labeled by the eigenvalues of $r$).
$$j={_{k+2}\over^{2\pi}}\left(\matrix{i\partial\phi+:\beta\gamma:&
\partial\beta-2i\beta\partial\phi-:\beta^2\gamma:\cr
\gamma&-i\partial\phi-:\beta\gamma:}\right)\eqno(15)$$
gives the $\widehat{L\hspace{0.04cm}sl(2)}$ currents and turns
the Fock spaces into the (Wakimoto)
$\widehat{L\hspace{0.04cm}sl(2)}$-modules. For $r=1,\dots,k+1$,
using the (regularized) powers of the screening charge integral
$$Q(x)=\int\limits_x^{x+2\pi}\gamma(s)\hspace{0.05cm}:{\rm e}^{2i\phi(s)}:
\hspace{0.05cm}ds\hspace{0.05cm},\eqno(16)$$
one obtains the following ($x$-independent) complex of
Fock space modules ($p\equiv k+2$)
$$\cdots\ \smash{\mathop{\rightarrow}\limits^{Q^r}}\ {\cal F}_{-r+2p}\
\smash{\mathop{\rightarrow}\limits^{Q^{p-r}}}\ \ {\cal F}_r\
\ \smash{\mathop{\rightarrow}\limits^{Q^r}}\ {\cal F}_{-r}\
\smash{\mathop{\rightarrow}\limits^{Q^{p-r}}}\ {\cal F}_{r-2p}\
\smash{\mathop{\rightarrow}\limits^{Q^r}}\ \cdots\eqno(17)$$
whose cohomology is concentrated at ${\cal F}_r$ and provides
the HW representation of $\widehat{L\hspace{0.04cm}sl(2)}$
of level $k$ and spin ${_1\over^2}(r-1)$ [19]. The
intertwining chiral fields are
$$h=\left(\matrix{\beta\hspace{0.05cm}:{\rm e}^{-i\phi}:\hspace{0.05cm}
Q&\ \beta\hspace{0.05cm}:{\rm e}^{-i\phi}:\cr
:{\rm e}^{-i\phi}:\hspace{0.05cm}Q&\ :{\rm e}^{-i\phi}:}
\right).\eqno(18)$$
Each raw of the matrix defines an operator between the complexes
(17) which projected onto their cohomology gives the component
with magnetic number $\pm 1/2$ of the spin $1/2$ primary field
of the WZW model [38]. The real fields of the WZW theory are
bilinear combinations of fields $h$ and their right-moving partners.

\section{Lattice free fields}

\ \ \ One may build the representations of the lattice Kac-Moody
algebra ${\cal K}_N$ using (deformed) free fields. Below,
$\ q^{1/2}\equiv{\rm e}^{\pi i/(2(k+2))}\ $ so that $q$ is a primitive
$2p$-th root of unity. We shall need lattice versions of the
$u(1)$ current and of the $\beta\gamma$ system.

\subsection{$u(1)$ current}

The lattice $u(1)$ algebra ${\cal U}_N$
is generated by invertible elements $\Theta(n)$,
$n\in{\bf Z}_N$, with the relations
$$\Theta(n)\hspace{0.05cm}\Theta(n+1)
=q^{1/2}\hspace{0.05cm}\Theta(n+1)
\hspace{0.05cm}\Theta(n)\eqno(19)$$
(all other generators commute). Think about $\Theta(n)$ as of a regularized
version of $\ \exp[{i\int_{2\pi n/N}^{2\pi(n+1)/N}\partial
\phi(s)\hspace{0.05cm}ds}]\hspace{0.07cm}$. $\ {\cal U}_N$ becomes
a $*$-algebra if we put $\Theta(n)^*=\Theta(n)^{-1}$.
\vskip 0.2cm

Consider, for $N$ odd, the representations of
$\hspace{0.05cm}{\cal U}_N$ acting in the space spanned by
orthonormal vectors $|\underline\alpha\rangle$ where
${\underline\alpha}\equiv(\alpha_0,\alpha_2,\dots,\alpha_{N-1})\hspace{0.05cm}$,
$\hspace{0.05cm}\alpha_{2n}\in{\bf Z}_{4p}\hspace{0.05cm}$,
$\hspace{0.05cm}\sum\alpha_{2n}=0$:
\begin{eqnarray*}
\Theta(2n)\hspace{0.08cm}
|{\underline\alpha}\rangle\ \ \ \ \ &=&z_{2n}\hspace{0.05cm}
q^{\alpha_{2n}/2}\hspace{0.05cm}
|{\underline\alpha}\rangle\ \ {\rm for}\ 2n<N-1,\\
\Theta(2n+1)\hspace{0.05cm}
|{\underline\alpha}\rangle&=&z_{2n+1}\hspace{0.05cm}|(\alpha_0,\dots,
\alpha_{2n}+1,\alpha_{2n+2}-1,\dots\alpha_{N-1})\rangle\hspace{0.05cm},\\
\Theta(N-1)\hspace{0.11cm}|{\underline\alpha}\rangle &=&
z_{N-1}\hspace{0.05cm}q^{\alpha_{N-1}/2+(N+1)/4}\hspace{0.05cm}
|(\alpha_0-1,\alpha_2,..\\
&&\hspace{4.8cm}\dots,\alpha_{N-3},\alpha_{N-1}+1)\rangle
\end{eqnarray*}
\vskip -0.2cm
\noindent where $|z_n|=1$.
\vskip 0.3cm

\noindent{\bf Proposition.} [39] {\it The above formulae
give irreducible $*$-representations
of $\hspace{0.05cm}{\cal U}_N$. They are
equivalent iff they correspond to the same
eigenvalues of the central elements $\Theta(n)^{4p}=z(n)^{4p}$ and
$\Pi\equiv q^{-1/2}\Theta\nobreak(\nobreak N\nobreak-\nobreak 1\nobreak)
\cdots\Theta(1)\Theta(0)=\prod z_n$. Every irreducible
$*$-representation of $\hspace{0.05cm}
{\cal U}_N$ is equivalent to one of the above.}
\vskip 0.3cm

Below, we shall study uniquely representations with $\Theta(n)^{4p}=1$
(we could have included this condition into the defining relations
of $\hspace{0.05cm}{\cal U}_N$). In those representations,
the rigid rotation of ${\bf Z}_N$ by two units may be implemented
by elements $U\in{\cal U}_N$:
$$U\hspace{0.05cm}\Theta(n+1)=\Theta(n-1)\hspace{0.05cm}U$$
where
$$U\hspace{0.05cm}=\hspace{0.05cm}(4p)^{-N/2}
\sum\limits_{\alpha_n\in{\bf Z}_{4p}}q^{-{_1\over^2}\alpha_0\alpha_{N-1}}
\hspace{0.05cm}\Theta(N-1)^{\alpha_{N-1}}\hspace{0.05cm}\cdots
\hspace{0.05cm}\Theta(1)^{\alpha_1}\hspace{0.05cm}\Theta(0)^{\alpha_0}.
\eqno(20)$$
Notice that, except for the boundary term, $U$ is a product
of local expressions. Eq. (20) gives the lattice version
of the Sugawara construction of the WZW Hamiltonian for the
abelian group.
\vskip 0.2cm

The case of $N$ even is similar.

\subsection{$\beta\gamma$ system}

\ \ \ Consider algebra ${\cal B}$ with generators ${\rm B}$,
$\Gamma$ and relation
$$q\hspace{0.05cm}{\rm B}\hspace{0.05cm}\Gamma-q^{-1}\hspace{0.05cm}
\Gamma\hspace{0.05cm}{\rm B}=q-q^{-1}.\eqno(21)$$
${\rm B}^p$ and $\Gamma^p$
generate the center of $\cal B$ and we have different classes
of representations depending on the eigenvalues of those elements.
\vskip 0.2cm

\begin{enumerate}
\item If in an irreducible representation
$\Gamma^p\not=0$ then $\Gamma$ is invertible and we may
introduce\footnote{I owe this observation to L. Faddeev and A. Volkov}
${\rm D}=\Gamma^{-1}-{\rm B}$ satisfying
$$\Gamma\hspace{0.05cm}{\rm D}=q^2\hspace{0.05cm}{\rm D}\hspace{0.05cm}
\Gamma.$$ If ${\rm D}=0$ we get the 1-dimensional representation\footnote{This
possibility was pointed to me by R. Kashaev}
of $\cal B$. If $\rm D\not=0$, the representations of $\cal B$ are
periodic: we may find a basis $|s\rangle$,
$s\in{\bf Z}_p$\hspace{0.05cm}, s.t.
$$\Gamma\hspace{0.05cm}|s\rangle=\zeta_1\hspace{0.05cm}|s+1\rangle,
\ \ \ {\rm D}\hspace{0.05cm}|s\rangle=\zeta_2\hspace{0.05cm}q^{-2s}
\hspace{0.05cm}|s\rangle.$$ The periodic representations are characterized
by the eigenvalues of $\hspace{0.05cm}\Gamma^p=\zeta_1^p$ and
of $\hspace{0.05cm}{\rm B}^p=\zeta_1^{-p}
+\zeta_2^p$.
\vskip 0.2cm
\item The case $\Gamma^p=0$ but ${\rm B}^p\not=0$ may be
treated similarly.
\vskip 0.2cm
\item Finally, if $\Gamma^p=0$ and ${\rm B}^p=0$, we either have
a trivial representation or a HW representation in $p$-dimensional
space spanned by states $|s\rangle$, $s=0,1,\dots,p-1,$ with the action
$$\Gamma\hspace{0.07cm}|s\rangle=
\cases{|s+1\rangle\ \ {\rm if}\ s<p-1,\hspace{1.06cm}\cr
0\ \ \ {\rm if}\ \ s=p-1,\hspace{1.06cm}}$$
$${\rm B}\hspace{0.05cm}|s\rangle=\cases{(1-q^{-2s})\hspace{0.05cm}
|s-1\rangle\ \ {\rm if}\ s>0,\cr \hspace{0.05cm}0\ \ \ {\rm if}\ \ s=0.}$$
\end{enumerate}
\vskip 0.2cm

The lattice $\beta\gamma$ system is obtained by taking a copy of algebra
$\cal B$ for each lattice site:
$${\cal B}_N=\bigotimes\limits_{n\in{\bf Z}_N}{\cal B}\hspace{0.05cm}.$$
We shall denote its generators as ${\rm B}(n)$ and $\Gamma(n)$.
${\cal B}_N$ may be represented in the space spanned by vectors
$\hspace{0.05cm}|{\underline s}\rangle\hspace{0.05cm}$, $\hspace{0.05cm}{\rm B}
(n)\hspace{0.05cm},\ \Gamma(n)\hspace{0.05cm}$ acting on the $n$-th
component of ${\underline s}\equiv(s_0,s_1,\dots,s_{N-1})\hspace{0.05cm}$.
Below, we shall only consider the HW representations of
${\cal B}_N$ so that we could include ${\rm B(n)}^p=\Gamma(n)^p=0$
into the defining relations of ${\cal B}_N$. The periodic
representations might be also of interest but we shall not
study them here.
\vskip 0.2cm
In the HW representation, the rigid rotation of ${\bf Z}_N$
by one unit is implemented by element $U\in{\cal B}_N$,
\begin{eqnarray*}
\hspace*{-0.39cm}U\hspace{0.05cm}=&&\hspace{-0.05cm}p^{-N}\hspace{-0.18cm}
\sum\limits_{\alpha_n,\hspace{0.03cm}\beta_n\in{\bf Z}_p}\
\prod\limits_{n=0}^{N-1}(1-q^{-2\beta_n})^{-1}
(1-q^{-2(\beta_n-1)})^{-1}\cdots(1-q^{-2})^{-1}\\
&&\cdot\hspace{0.05cm}\Gamma(N-1)^{\beta_0}
\hspace{0.05cm}(1-\Gamma(N-1){\rm B}(N-1))^{\alpha_{N-1}}
\hspace{0.05cm}{\rm B}(N-1)^{\beta_{N-1}}\cdots\cdots\cdot\\
&&\cdots\hspace{0.05cm}\Gamma(1)^{\beta_2}\hspace{0.05cm}
(1-\Gamma(1){\rm B}(1))^{\alpha_1}\hspace{0.05cm}{\rm B}(1)^{\beta_1}
\hspace{0.05cm}\Gamma(0)^{\beta_1}\hspace{0.05cm}
(1-\Gamma(0){\rm B}(0))^{\alpha_0}\hspace{0.05cm}{\rm B}(0)^{\beta_0}
\hspace{0.05cm},
\end{eqnarray*}
such that
\begin{eqnarray*}
U\hspace{0.05cm}{\rm B}(n+1)&=&{\rm B}(n)\hspace{0.05cm}U\hspace{0.05cm},\\
U\hspace{0.05cm}\Gamma(n+1)&=&\Gamma(n)\hspace{0.05cm}U\hspace{0.05cm}.
\end{eqnarray*}

\subsection{Lattice Wakimoto representation}

\ \ \ It will be convenient to put together the lattice $u(1)$ fields
and the $\beta\gamma$ system in a somewhat twisted way defining their
action on vectors $\hspace{0.05cm}|{\underline\alpha},{\underline s}\rangle
\hspace{0.05cm}\equiv|{\underline\alpha}\rangle\otimes|{\underline s}\rangle
\hspace{0.05cm}$ by
\begin{eqnarray*}
\Theta(n)\hspace{0.05cm}|{\underline\alpha},{\underline s}\rangle
&=&q^{s_{n+1}-s_n}\hspace{0.05cm}(\Theta(n)\hspace{0.05cm}|
{\underline\alpha}\rangle)\otimes|{\underline s}\rangle\hspace{0.05cm},\\
{\rm B}(n)\hspace{0.05cm}|{\underline\alpha},{\underline s}\rangle
&=&|{\underline\alpha}\rangle\otimes({\rm B}(n)\hspace{0.05cm}
|{\underline s}\rangle)\hspace{0.05cm},\\
\Gamma(n)\hspace{0.05cm}|{\underline\alpha},{\underline s}\rangle
&=&|{\underline\alpha}\rangle\otimes(\Gamma(n)\hspace{0.05cm}
|{\underline s}\rangle)\hspace{0.05cm}.
\end{eqnarray*}
Taking different representations of $\hspace{0.05cm}{\cal U}_N
\hspace{0.05cm}$ and the HW representation of
$\hspace{0.05cm}{\cal B}_N$, we obtain this way irreducible representations
(labelled by the eigenvalue $z$ of $\Pi$) of algebra $\hspace{0.05cm}{\cal U}_N
\tilde{\otimes}{\cal B}_N\hspace{0.05cm}$, \hspace{0.05cm}a twisted
tensor product of $\hspace{0.1cm}{\cal U}_N\hspace{0.05cm}$
and $\hspace{0.05cm}{\cal B}_N\hspace{0.05cm}$. We shall denote
by ${\cal H}_z$ the corresponding representation space.
\vskip 0.2cm
\addtocounter{equation}{21}
\renewcommand{\theequation}{\arabic{equation}}

The lattice version of the Wakimoto realization (15) of the Kac-Moody
currents by free fields is given by the formulae
\begin{eqnarray}
J_{11}(n)&=&\Theta(n)+q^{-1/2}\Theta(n)^{-1}{\rm B}(n+1)
\Gamma(n)\hspace{0.05cm},\cr
J_{12}(n)&=&-\Theta(n){\rm B}(n)+q^{-1/2}\Theta(n)^{-1}{\rm B}(n+1)
(1-\Gamma(n){\rm B}(n))\hspace{0.05 cm},\\
J_{21}(n)&=&q^{1/2}\Theta(n)^{-1}\Gamma(n)\hspace{0.05cm},\cr
J_{22}(n)&=&q^{1/2}\Theta(n)^{-1}(1-\Gamma(n){\rm B}(n))
\nonumber
\end{eqnarray}
which define a homomorphism from the lattice Kac-Moody algebra
$\hspace{0.05 cm}{\cal K}_N\hspace{0.05 cm}$ to
$\hspace{0.05 cm}{\cal U}_N\tilde{\otimes}{\cal B}_N\hspace{0.05 cm}$
and turn each representation space ${\cal H}_z$ of the latter into a
(Wakimoto) ${\cal K}_N$-module. These modules are irreducible
if $z^{2p}\not=1$. For $z=q^r$, $r=-p+1,\dots,p-1,p$, their reducibility
may be studied by adapting to the lattice the cohomological constructions
of [19].

\section{Bernard-Felder cohomology}

We shall have to adjoin to the $u(1)$ algebra $\hspace{0.05cm}{\cal U}_N$
the zero mode $\Psi(0)$ s.t.
\begin{eqnarray*}
\hbox to 4.5cm{\hspace{2cm}$\Theta(0)\hspace{0.05cm}\Psi(0)$\hfill}&=&
q^{-1/2}\hspace{0.05cm}\Psi(0)\hspace{0.05cm}\Theta(0)\hspace{0.05cm},\cr
\hbox to 4.5cm{\hspace{2cm}$\Theta(N-1)\hspace{0.05cm}\Psi(0)$\hfill}&=&
q^{-1/2}\hspace{0.05cm}\Psi(0)\hspace{0.05cm}\Theta(N-1)
\end{eqnarray*}
and all other commutators are trivial. $\Psi(0)$, which divides the
eigenvalue of $\Pi$ by $q$,  may be implemented
in the sum of representation spaces of $\hspace{0.05cm}{\cal U}_N$
with $\Pi=q^r$. We shall also let it act in $\oplus {\cal H}_{q^r}$ by
$$\Psi(0)\hspace{0.05cm}|{\underline\alpha},{\underline s}\rangle
=q^{s_0}\hspace{0.05cm}(\Psi(0)\hspace{0.05cm}|{\underline\alpha}\rangle)
\otimes|{\underline s}\rangle\hspace{0.05cm}.$$
More symmetrically, we may construct the lattice $u(1)$ vertex
operator ($\sim$ ${\rm e}^{i\phi(2\pi n/N)}$)
\begin{eqnarray}
\Psi(n)&=&\Theta(n-1)\hspace{0.05cm}...\hspace{0.05cm}
\Theta(1)\hspace{0.05cm}\Theta(0)\hspace{0.05cm}\Psi(0)\ \ \ \ \ \ \ \ \
\ \ \ {\rm for}\ n>0\hspace{0.05cm},\cr
\Psi(n)&=&\Theta(n)^{-1}...\hspace{0.05cm}\Theta(-2)^{-1}
\Theta(-1)^{-1}\Psi(0)\ \ \ \ {\rm for}\ n<0\ .
\end{eqnarray}
The screening charge integral is now defined by
$$Q(n)=\Pi^{-1}\sum\limits_{m=n}^{n+N-1}\Gamma(m)\Psi(m)^2
\hspace{0.05cm},\eqno(24)$$
compare eq.\hspace{-0.05cm} (16). It is related to the lower left
matrix element of the monodromy matrix (10) in the Wakimoto
realization (22):
$$Q(n)=q\hspace{0.05cm}M_{21}(n)\hspace{0.05cm}\Psi(n)^2\hspace{0.05cm}$$
and thus (see eq.\hspace{-0.05cm} (11))
to the $\hspace{0.05cm}{\cal U}_q(sl(2)$ lowering operator. The relation of
the screening charge integrals to quantum groups has been observed
in [20] and was developed in [21]-[23] into a theory of topological
realizations of quantum groups.
\vskip 0.2cm

$Q(n)$ acts as an operator from ${\cal H}_{z}$ to ${\cal H}_{z'}$,
$z'=q^{-2}z$,
in a nilpotent
way: $Q(n)^p=0$. Besides, for $r=0,1,\dots,p$\hspace{0.05cm},
powers of $Q(n)$ define complexes of ${\cal K}_N$-modules
\begin{eqnarray*}
&\hspace{1.87cm}0\ \rightarrow\ {\cal H}_{q^{-r}}\
\smash{\mathop{\rightarrow}\limits^{Q^{p-r}}}
\ {\cal H}_{q^{r}}\ \smash{\mathop{\rightarrow}
\limits^{Q^{r}}}\ {\cal H}_{q^{-r}}\ \rightarrow\ 0&\ ,
\end{eqnarray*}
\vskip -0.4cm
\begin{eqnarray*}
&\hspace{2cm}0\ \rightarrow\ {\cal H}_{q^r}\
\smash{\mathop{\rightarrow}\limits^{Q^{r}}}
\ {\cal H}_{q^{-r}}\ \smash{\mathop{\rightarrow}\limits^{Q^{p-r}}}\ {\cal
H}_{q^r}\
\rightarrow\ 0&\ .
\end{eqnarray*}
In other words, the powers of $Q(n)$ above (independent, in fact, of $n$)
commute with the action of ${\cal K}_N$ as given by eqs.\hspace{-0.05cm}
(22). One may show that the above complexes are exact in the middle.
We conjecture that the remaining cohomology
\addtocounter{equation}{1}
\renewcommand{\theequation}{\arabic{equation}}
\begin{eqnarray}
\hspace{1cm}\ \ \ {\cal H}_{q^r}\supset
{\rm ker}\hspace{0.05cm}\hspace{0.05cm} Q^r\hspace{0.05cm}\equiv
{\cal H}'_{q^r}\hspace{0.05cm}
\ \cong\ \hspace{0.05cm}{\cal H}_{q^{-r}}\hspace{-0.05cm}/\hspace{0.05cm}
\hspace{0.05cm}{\rm im}\hspace{0.05cm}\hspace{0.05cm} Q^{r}\hspace{0.05cm}
\equiv\hspace{0.05cm}{\cal H}'_{q^{-r}}
\end{eqnarray}
\vskip -0.2cm
\noindent and
\vskip -0.7 cm
\begin{eqnarray}
\hspace{1cm}{\cal H}_{q^{-r}}\supset
{\rm ker}\hspace{0.05cm}\hspace{0.05cm} Q^{p-r}\hspace{0.05cm}
\equiv\hspace{0.05cm}{\cal H}''_{q^{-r}}\hspace{0.05cm}
\ \cong\ \hspace{0.05cm}{\cal H}_{q^{r}}\hspace{-0.05cm}/\hspace{0.05cm}
\hspace{0.05cm}{\rm im}\hspace{0.05cm}\hspace{0.05cm} Q^{p-r}
\hspace{0.05cm}\equiv\hspace{0.05cm}{\cal H}''_{q^r}
\end{eqnarray}
gives irreducible representations of ${\cal K}_N$.
This is true, for example, for $N=1$ when ${\cal K}_N$ reduces to
$\hspace{0.05cm}{\cal U}_q(sl(2))$\hspace{0.05cm}.
\vskip 0.2cm

\section{WZW theory on lattice}

\ \ \ The chiral intertwining operators on the lattice are
$$h(n)=\left(\matrix{(\hspace{0.05cm}\Pi-\Pi^{-1})\hspace{0.05cm}\Psi(n)+
{\rm B}(n)\hspace{0.05cm}\Psi(n)^{-1}\hspace{0.05cm} Q(n)&
\ {\rm B}(n)\hspace{0.05cm}\Psi(n)^{-1}\cr
\Psi(n)^{-1}\hspace{0.05cm} Q(n)&\ \Psi(n)^{-1}}\right).\eqno(27)$$
In comparison with the expression (18), notice additional
term in the uper left corner of (27) killed by Wick ordering renormalization
in the continuum.
Intertwiners (27) map space $\oplus {\cal H}_{q^r}$ into itself and descend
to cohomology (25),\hspace{0.05cm}(26).
\vskip 0.2cm

The above constructions have been adapted to the left-moving sector
of the WZW theory. In the ones for the right-moving sector (distinguished
below by the bar), we should
replace $q$ by $\bar q=q^{-1}$. The space of states of the lattice WZW theory
may be taken as
$${\cal H}'\ =\ \bigoplus\limits_{r=1}^{p-1}{\cal H}'_{q^r}
\otimes\bar{\cal H}'_{q^r}$$
or as ${\cal H}''$ using the doubly-primed spaces. The real fields
of the lattice WZW model are then
$$g(n,m)=h(n)\hspace{0.05cm}\bar h(m)^{-1}\eqno(28)$$
acting in spaces ${\cal H}'\ {\rm or}\ {\cal H}''$, $\hspace{0.05cm}
g(n,m)=g(n+N,m+M)$.
\vskip 0.2cm

It should be stressed that the resulting theory lives
on Minkowski lattice ($n,\hspace{0.05cm}m$ in eq.\hspace{-0.05cm}
(28) are integer valued light-cone variables). The model has
essentially all the symmetries of the continuum WZW theory (in deformed form),
including the conformal covariance. Its relation to quantum groups
is explicit. Let us conclude by listing some
open problems.
\vskip 0.2cm

\begin{enumerate}
\item As mentioned above, the lattice counterpart of the Sugawara
construction of energy-momentum is not known apart from the abelian
case.
\vskip 0.1cm

\item The continuum limit is not easy to understand even for the
$u(1)$ case. One should expect it to take place in a weak form,
for vacuum expectation values or traces of products of operators.
Studying the first ones would require a choice of vacuum on
the lattice (among many states invariant under lattice translations).
The traces (e.g. the characters of the lattice Kac-Moody algebra)
seem more accessible.
\vskip 0.1cm

\item In the above constructions we did not study the unitarity
properties of representations of ${\cal K}_N$ or of the
intertwining fields. One knows that in the continuum this
is difficult within the free field approach, even
when the cohomology of the Fock space modules gives
unitary HW representations of the Kac-Moody algebra.
It remains to be seen if there exists a (deformed?) version
of unitarity properties on the lattice.
\vskip 0.1cm

\item There is a close relation between the two-dimensional
(chiral) WZW model and the three dimensional Chern-Simons theory [40].
For example, the space of the fixed-time Chern-Simons states on the disc
is the basic representation of the Kac-Moody algebra [41].
The Chern-Simons theory seems to possess lattice
versions given by constructions [42][43] of 3-manifold invariants
from quantum groups. The lattice WZW model described above should
be related to the latter. It seems that the clue to understanding
this relation should be the fusion of lattice Kac-Moody algebras
generalizing the coproduct for quantum groups and allowing to glue
spaces of Chern-Simons states for more complicated topologies.
\end{enumerate}


\begin{thebibliography}{}

\bibitem{} E. Witten, Non-Abelian Bosonization in Two Dimensions, {\it
Commun. Math. Phys.}, {\bf 92}, 455-472 (1984).

\bibitem{} P. Goddard, A. Kent, D. Olive, Unitary Representations of
the Virasoro and Super-Virasoro Algebras, {\it Commun. Math. Phys.},
{\bf 103}, 105-119 (1986).

\bibitem{} V. Drinfeld, V. Sokolov, Lie Algebras and Equations of
Korteweg-de Vries type, {\it Journ. Sov. Math.}, {\bf 30}, 1975-2036 (1985).

\bibitem{} O. Babelon, Extended Conformal Algebra and Yang-Baxter Equation,
{\it Phys. Lett.}, {\bf B 215}, 523-529 (1988).

\bibitem{} G. Felder, J. Fr\"{o}hlich, G. Keller, Braid Matrices
and Structure Constants for Minimal Conformal Models,
{\it Commun. Math. Phys.}, {\bf 124}, 647-664 (1989).

\bibitem{} J.-L. Gervais, The Quantum Group Structure of 2d Gravity and Minimal
Models I, {\it Commun. Math. Phys.}, {\bf 130}, 257-283 (1990).

\bibitem{} K. Fredenhagen, K. Rehren, B. Schroer, Superselection
Sectors with Braid Group Statistics and Exchange Algebras,
{\it Commun. Math. Phys.}, {\bf 125}, 201-226 (1989)

\bibitem{} J. Fr\"{o}hlich, F. Gambbiani, Braid Statistics
in Local Quantum Theory, ETH-TH/90-6 preprint.

\bibitem{} G. Mack, V. Schomerus, Conformal Field Algebras
with Quantum Symmetry from the Theory of Superselection Sectors,
{\it Commun. Math. Phys.}, {\bf 134}, 139-196 (1990)

\bibitem{} S. Doplicher, R. Haag, J. Roberts, Local
Observables  and Particle Statistics I and II, {\it Commun.
Math. Phys.}, {\bf 23}, 199-230 (1971) and {\bf 35}, 49-85 (1974).

\bibitem{} S. Doplicher, J. Roberts, Why There is a Field Algebra
with a Compact Gauge Group Describing the Superselection Structure
in Particle Physics, {\it Commun. Math. Phys.}, {\bf 131}, 51-108
(1990)

\bibitem{} V. Pasquier, H. Saleur, Common structures between finite systems
and conformal field theories through quantum groups,
{\it Nucl. Phys.}, {\bf B 330}, 523-556 (1990).

\bibitem{} V. Drinfeld, Quantum Groups, in Proceedings of the
International Congress of
Mathematicians, Berkeley 1986, pp. 798-820.

\bibitem{} M. Jimbo, A $q$-difference Analogue of $U(g)$ and Yang-Baxter
Equation,
Lett. Math. Phys., {\bf 10}, 63-69 (1985).

\bibitem{} T. Kohno, Monodromy representations of braid groups and Yang-Baxter
equations, {\it Ann. Inst. Fourier (Grenoble)}, {\bf 37(4)}, 139-160 (1987).

\bibitem{} G. Moore, N. Seiberg, Classical and Quantum Conformal Field
Theory, {\it Commun. Math. Phys.}, {\bf 123}, 177-254 (1989).

\bibitem{} L. Alvarez-Gaum\'{e}, C. G\'{o}mez, G. Sierra, Duality and
Quantum Groups, {\it Nucl. Phys.}, {\bf B 330}, 347-398 (1990)

\bibitem{} M. Wakimoto, Fock Representations of the Affine Lie Algebra
$A^{(1)}_1$,
{\it Commun. Math. Phys.}, {\bf 104} 605-609 (1986).

\bibitem{} D. Bernard, G. Felder, Fock Representations
and BRST Cohomology in $SL(2)$ Current Algebra, {\it Commun.
Math. Phys.}, {\bf 127}, 145-168 (1990).

\bibitem{} C. G\'{o}mez, G. Sierra, Quantum Group Meaning of the Coulomb Gas,
{\it Phys. Lett.}, {\bf B 240}, 149-157 (1990).

\bibitem{} V. Schechtman, A. Varchenko, Arrangements of Hyperplanes
and Lie Algebra Homology, {\it Invent. Math.}, to appear.

\bibitem{} R. Lawrence, The Topological Approach to Representations
of Iwahori-Hecke Algebra, {\it Int. J. Mod. Phys.}, {\bf A 5},
3213-3219 (1990).

\bibitem{} G. Felder, C. Wieczerkowski, Topological Representations
of th Quantum Group $U_q(sl_2)$, {\it Commun. Math. Phys.},
{\bf 138}, 583-605 (1991).

\bibitem{} E. Verlinde, Fusion Rules and Modular Transformations
in 2d Conformal Field Theory, {\it Nucl. Phys.}, {\bf{B\ 300}} [FS\ 22],
360-376 (1988).

\bibitem{} G. Moore, N. Reshetikhin, A Comment on Quantum Group Symmetry
in Conformal Field Theory, {\it Nucl. Phys.}, {\bf B 328}, 557-574 (1989).

\bibitem{} A. Alekseev, L. Faddeev and M. Semenov-Tian-Shansky,
Hidden quantum groups inside Kac-Moody algebras, in Quantum Groups,
ed. P. Kulish, Lect. Notes in Math. Vol. 1510,, Berlin, Springer 1992,
pp. 148-158.

\bibitem{} A. Alekseev, L. Faddeev, M. Semenov-Tian-Shansky and A. Volkov,
The unraveling of the quantum group structure in the WZNW theory,
CERN preprint 1991.

\bibitem{} L. Freidel, J.M. Maillet, On Classical and Quantum Integrable
Field Theories Associated to Kac-Moody Current Algebras,
{\it Phys. Lett.}, {\bf {B 263}}, 403-410 (1991).

\bibitem{} L. Faddeev, Quantum Symmetry in Conformal Field Theory
by Hamiltonian Methods, Helsinki preprint 1992.

\bibitem{} A. Volkov, Miura Transformation on the Lattice,
{\it Theor. Math. Phys.}, {\bf 74}, 135-139 (1988).

\bibitem{} O. Babelon, Exchange Formula and Lattice Deformation
of the Virasoro Algebra, {\it Phys. Lett.}, {\bf B 238}, 234-238 (1990).

\bibitem{} O. Babelon, L. Bonora, Quantum Toda Theory,
{\it Phys. Lett.}, {\bf B 253}, 365-372 (1991).

\bibitem{} L. Bonora, V. Bonservizi, Quantum $sl_n$ Toda field theories,
SISSA-ISAS preprint 1992.

\bibitem{} V. Kac, Infinite Dimensional Lie Algebras, Cambridge,
Cambridge University Press 1985.

\bibitem{} A. Pressley, G. Segal, Loop Groups, Oxford, Clarendon Press 1986.

\bibitem{} B. Feigin, E. Frenkel, Representations of Affine Lie Kac-Moody
algebras and bosonization, in V.G. Knizhnik Memorial Volume,
Singapore, World Scientific 1989.

\bibitem{} P. Bouwknegt, J. McCarthy, K. Pilch, Quantum Group Structure
in the Fock Space Resolutions of $\hspace{0.15cm}
\hat{}\hspace{-0.15cm}sl(n)$ Representations,
{\it Commun. Math. Phys.}, {\bf 131}, 125-155 (1990).

\bibitem{} A. Tsuchiya, Y. Kanie, Vertex Operators in the Conformal
Field Theory on P1 and Monodromy Representations of the Braid
Group, {\it Adv. Stud. Pure Math.}, {\bf 16}, 297-372 (1988).

\bibitem{} K. Gaw\c{e}dzki, On Lattice Kac-Moody Algebras,
in preparation.

\bibitem{} E. Witten, Quantum Field Theory and Jones Polynomial,
{\it Commun. Math. Phys.}, {\bf{121}}, 351-399 (1989).

\bibitem{} S. Elitzur, G. Moore, A. Schwimmer, N. Seiberg, Remarks on the
Canonical Quantization of the Chern-Simons-Witten Theory, {\it Nucl. Phys.},
{\bf{B\ 326}}, 104-134 (1989).

\bibitem{} N. Reshetikhin, V. Turaev, Invariants of 3-Manifolds
via Link Polynomials and Quantum Groups, {\it Inven. Math.},
{\bf 103}, 547-597 (1991).

\bibitem{} V. Turaev, O. Viro, State sum of 3-manifolds and quantum
6j-symbols, LOMI preprint 1990.



\end{thebibliography}
\end{document}